\pdfoutput=1
\documentclass[superscriptaddress,showpacs,aps,prd,twocolumn]{revtex4-1}

\usepackage[colorlinks=true, pdfstartview=FitV, linkcolor=red, citecolor=blue,
urlcolor=blue]{hyperref}
\usepackage[dvipdfmx]{graphicx}
\usepackage{amsmath,amssymb}
\usepackage{color}

\newcommand{\Lag}{\mathcal{L}}

\newcommand{\CP}{\mathcal{CP}}
\newcommand{\calC}{\mathcal{C}}
\newcommand{\ls}{l_\text{system}}
\newcommand{\tr}{\text{tr}}

\newcommand{\bB}{{\boldsymbol B}}

\newcommand{\bsigmap}{{\boldsymbol \sigma}_\perp}
\newcommand{\bDp}{{\boldsymbol D}_\perp}
\newcommand{\hg}{{}_1F_1}
\newcommand{\pp}{p_\perp}
\newcommand{\plk}{p_{l,k}}

\newcommand{\llk}{\lambda_{l,k}}

\newcommand{\philk}{\phi_{l,k}}
\newcommand{\varphilk}{\varphi_{l,k}}
\newcommand{\tphilk}{\widetilde{\phi}_{l,k}}
\newcommand{\tvarphilk}{\widetilde{\varphi}_{l,k}}
\newcommand{\Nlk}{N_{l,k}}

\begin{document}

\title{Surface magnetic catalysis}

\author{Hao-Lei~Chen}
\affiliation{Physics Department and Center for Particle Physics and
             Field Theory, Fudan University, Shanghai 200433, China}
\author{Kenji~Fukushima}
\affiliation{Department of Physics, The University of Tokyo,
             Tokyo 113-0033, Japan}
\author{Xu-Guang~Huang}
\affiliation{Physics Department and Center for Particle Physics and
             Field Theory, Fudan University, Shanghai 200433, China}
\affiliation{Key Laboratory of Nuclear Physics and Ion-beam Application (MOE), 
             Fudan University, Shanghai 200433, China}             
\author{Kazuya~Mameda}
\affiliation{Physics Department and Center for Particle Physics and
             Field Theory, Fudan University, Shanghai 200433, China}
\affiliation{Department of Physics, The University of Tokyo,
             Tokyo 113-0033, Japan}
                                       
\begin{abstract}
We study fermions in a magnetic field in a finite-size cylinder.
With the boundary condition for the fermion flux, we show that the energy spectra and the wave functions are modified by the finite-size effect;
the boundary makes the degenerate Landau levels appear only partially for states with small angular momenta, while the boundary effect becomes stronger for states with large angular momenta.
We find that mode accumulation at the boundary occurs for large angular momenta and that the magnetic effect is enhanced on the boundary surface.
Using a simple fermionic model, we quantify the magnetic catalysis, i.e.\ the magnetic enhancement of the fermion pair condensation, in a finite-size cylinder.
We confirm that the magnetic catalysis is strongly amplified at the boundary due to the mode accumulation.
\end{abstract}


\maketitle

\section{Introduction}

Magnetic field backgrounds add many intriguing aspects in quantum
many-body systems.  In quantum chromodynamics (QCD),  theoretical
studies of significant interest in magnetic responses have been
inspired by gigantic magnetic fields which could exist in the early
Universe~\cite{Cheng:1994yr,*Baym:1995fk,*Grasso:2000wj}, compact
stars~\cite{Duncan:1992hi}, and relativistic heavy-ion
collisions~\cite{Skokov:2009qp,*Voronyuk:2011jd,*Deng:2012pc}.  One
recent and actively discussed example is the anomalous transport
phenomenon, such as the chiral magnetic effect and its
relatives~\cite{Fukushima:2008xe,*Son:2009tf,*Kharzeev:2015znc,*Huang:2015oca},
in a quark-gluon plasma involving an external magnetic field.

The QCD vacuum structure is also quite sensitive to the magnetic
field;  a pair condensate of fermions and antifermions or the chiral
condensate is enhanced by the magnetic field, which is called the
magnetic
catalysis~\cite{Klimenko:1991he,*Klimenko:1992ch,*Gusynin:1994re,*Gusynin:1995nb}.
This well-known feature in a magnetic field applied to QCD was
originally obtained in the framework of the Nambu--Jona-Lasinio (NJL)
model.  Since then, the magnetic catalysis has been theoretically
investigated with various models and various approaches: the
quark-meson
model~\cite{Fraga:2008qn,*Mizher:2010zb,*Andersen:2011ip,*Ferrari:2012yw},
the MIT bag model~\cite{Fraga:2012fs}, the lattice QCD
simulation~\cite{Bali:2011qj,*Bali:2012zg,*Bruckmann:2013oba}, the
holographic model~\cite{Johnson:2008vna} and the renormalization group
analysis~\cite{Hong:1996pv,*Skokov:2011ib,*Scherer:2012nn,*Andersen:2012bq,*Fukushima:2012xw,*Kamikado:2013pya,*Hattori:2017qio}
(see also Ref.~\cite{Shovkovy:2012zn,*Miransky:2015ava} for recent
reviews and the references therein).
Fascinatingly, some nontrivial interplay between other external
influences and the magnetic field leads to more subtle changes in the
QCD vacuum.  Contrary to what is expected from the magnetic catalysis,
a strong magnetic field can melt the chiral condensate and restore a
part of broken chiral symmetry once the magnetic field is coupled with
finite-density and finite-temperature effects, which are called inverse magnetic
catalysis~\cite{Ebert:1999ht,*Preis:2010cq,*Preis:2012fh} or magnetic inhibition~\cite{Fukushima:2012kc}.
The rich structure of the QCD vacuum influenced by the magnetic field
is also discussed in a globally rotating system~\cite{Chen:2015hfc}.

The robustness for the abovementioned magnetic phenomena is ensured
by the characteristic energy spectrum of charged particles in the
magnetic field, namely, the Landau quantization with discrete Landau
levels;  for fermions with charge $e$ in the external magnetic field $B$, 
the transverse momenta perpendicular
to the magnetic field are replaced by $\pp=\sqrt{eB(2n+1-2s_z)}$,
with $n=0,1,\dots$ and $s_z=\pm\tfrac{1}{2}$.  In other
words, once the pattern of the Landau quantization is distorted, a
novel aspect of magnetic QCD dynamics may be expected.  In real
physical systems which have a finite-size, such a modification
inevitably appears through the boundary condition.  We thus expect the boundary condition to affect the energy dispersion generally in
matter under the magnetic field, which is understood from the
following argument.
The length scale of the cyclotron motion (i.e.\ the Larmor radius) is
characterized by the magnetic length, $l_B\equiv 1/\sqrt{eB}$.  As
long as the length scale of the system, which is denoted by $\ls$, is
much larger than $l_B$, particles do not feel the presence of the
boundary.  In this case, corresponding to the quantized cyclotron
motion, the well-known conventional Landau levels $\sim 1/l_B$ are
formed.  By contrast, for $\ls\lesssim l_B$, the cyclotron motion with
a large radius is disturbed by the boundary, and thus the ordinary
Landau quantized spectra are no longer obtained.  Specifically, in
the weak magnetic field limit, the transverse momenta should be of
order not $\sim 1/l_B$ but $\sim 1/\ls$.

On top of the fact that real physical systems have a finite size, we
have a strong motivation to formulate the finite-size effect for a
rotating system, e.g.\ a rotating quark-gluon plasma whose orbital
angular momentum is provided by the noncentral geometry in the
relativistic heavy-ion
collision~\cite{Liang:2004ph,*Huang:2011ru,*Becattini:2015ko,*Jiang:2016woz,*Aristova:2016wxe,*Deng:2016gyh}.
Let us briefly review the finite-size effect on rotating matter for
$B=0$.  For a rotating system, it is crucially important to impose a boundary
condition at a finite distance from the rotational center;  otherwise,
the speed of the rotational motion exceeds the speed of the light and
the relativistic causality is violated.
If we impose a proper boundary condition, we can verify that uniform rotation
alone would not affect the vacuum structure because all excitations
are gapped~\cite{Davies:1996ks,Ambrus:2015lfr}.  Hence, we can say
that, at zero temperature without any other external source, the
rotational effect on fermionic thermodynamics is
invisible~\cite{Ebihara:2016fwa}.  At finite temperature or density,
on the other hand, the chiral phase transition feels the effective
chemical potential induced by
rotation~\cite{Jiang:2016wvv,Chernodub:2016kxh,*Chernodub:2017ref,*Chernodub:2017mvp}.

From the above arguments, it would be expected that a finite-size
system with $B\neq 0$ should have complicated and interesting effects
which compete with each other.  One is the finite energy gap from the
boundary effect, and the other is the partial realization of the
gapless Landau zero modes.  In fact, unlike the rotation without any
other external source, a finite $B$ can change the vacuum of rotating
matter.  In Ref.~\cite{Chen:2015hfc}, the present authors first
discussed the low-energy fermionic dynamics under the presence of
finite magnetic field and rotation, and the authors showed that the
rotational effect leads to an inverse magnetic catalysis in the same
way as the finite density situation.  Also, an anomalous phenomenon in
the presence of vorticity (i.e.\ local rotation) and magnetic
background has been revealed in the formulation of
hydrodynamics~\cite{Hattori:2016njk} and quantum field
theory~\cite{Ebihara:2016fwa}.  In these analyses, however, only the
limit of $\ls\gg l_B$ was implicitly assumed (for simplicity).  For such
a large system, the angular velocity must be smaller than the system
size inverse in order not to violate the causality constraint.  Hence, we need
to consider the finite-size effect properly to make a theoretical
suggestion for thermodynamic properties of matter involving rapid
rotation (or large vorticity) coupled with the magnetic field.

In this paper we do not treat rotation but instead study a finite-size
cylindrical system under the magnetic field.  Although the coupling
with rotation is an important extension, we will see that the boundary
condition induces a highly nontrivial surface effect.  Imposing a
boundary condition for fermions, we numerically compute the energy
spectra and the wave functions of fermions at a finite $B$.  Then we
find that the Landau levels with a larger angular momentum are more
modified by the finite-size effect; that is, we observe incomplete or
nondegenerate Landau levels.  More importantly, we point out that the
mode accumulation occurs at the boundary surface.  For a concrete
demonstration with the NJL model in which obtained spectra and
wave functions are implemented, we calculate the chiral condensate or
the dynamical mass which is spatially dependent in finite-size
systems.  We then conclude that there emerges peculiar behavior of the
dynamical mass near the surface, which arises from the mode
accumulation there.  We call this novel phenomenon \textit{surface magnetic catalysis} in this work.

\section{Dirac equation with boundary}
\label{sec:Dirac}

We start our discussion with the Dirac equation under an external
constant magnetic field in systems with a finite size.  We choose the
magnetic field direction along the $z$ axis, i.e., $\bB=B\hat{z}$, and
we take the symmetric gauge with $A^\mu = (0,-By/2,Bx/2,0)$.  Then 
the Dirac equation reads
\begin{equation}
  \label{eq:Deq}
  \begin{split}
  \Bigl[i\gamma^0\partial_0 & + i\gamma^1(\partial_1 + ieBy/2) \\
  & + i\gamma^2(\partial_2 - ieBx/2)
    + i\gamma^3\partial_3 - m \Bigr]\psi = 0 \,.
  \end{split}
\end{equation}
Let us solve the above Dirac equation explicitly in the cylindrical
coordinates, $(t,r,\theta,z)$, with a boundary set at $r=R$.  In the
Dirac representation for $\gamma^\mu$'s, we write down two independent
positive-energy solutions with different spin polarizations but the
same total angular momentum $j=l+1/2$ (along the $z$ axis) as follows:
\begin{align}
\label{eq:u+}
  \psi = u_+ &= \frac{e^{-i\varepsilon t+ip_zz}}{\sqrt{\varepsilon+m}}
  \begin{pmatrix}
    (\varepsilon+m) \philk \\[1pt]
    0 \\[1pt]
    p_z \philk \\[3pt]
    i\sqrt{2eB\llk} \varphilk
  \end{pmatrix} ,\\
\label{eq:u-}
  \psi = u_- &= \frac{e^{-i\varepsilon t+ip_zz}}{\sqrt{\varepsilon+m}}
  \begin{pmatrix}
    0 \\[1pt]
    (\varepsilon+m) \varphilk \\[3pt]
    -i\sqrt{2eB\llk} \philk \\[1pt]
    -p_z \varphilk
  \end{pmatrix}
\end{align}
with $\varepsilon = \sqrt{2eB\llk+p_z^2+m^2}$.  Here, $\llk$
represents a modified Landau level index in a finite-size system and
its explicit form depends on the boundary condition at $r=R$.  We will
elucidate how to fix $\llk$ in the next section.  For the above
wave functions, we introduced a compact notation as
\begin{equation}
  \begin{split}
    & \philk \equiv e^{il\theta}\Phi_l (\llk,\tfrac{1}{2}eB r^2) \,,\\
    & \varphilk \equiv e^{i(l+1)\theta}\Phi_{l+1}(\llk-1,\tfrac{1}{2}eB r^2) \,.
  \end{split}
\label{eq:phis}
\end{equation}
We note that the above functions correspond to $\phi_\ell$ and
$\varphi_\ell$ in Ref.~\cite{Ebihara:2016fwa}.  The Landau
wave function is deformed by the finite-size effect, and for $l\ge 0$,
we find
\begin{equation}
  \begin{split}
  \Phi_{l\ge 0} (\lambda,x) &= \frac{1}{\Gamma(l+1)} \sqrt{
    \frac{\Gamma(\lambda+l+1)}{\Gamma(\lambda+1)} } \\
   &\qquad \times x^{\frac{l}{2}} e^{-x/2} \hg\bigl(-\lambda,l+1,x\bigr)\,.
  \end{split}
  \label{eq:Phi1}
\end{equation}
Here, $\hg(a,b,x)$ denotes the confluent hypergeometric function also as known as
Kummer's function of the first kind.  We chose the normalization to
recover the conventional spinors in the $R\to \infty$ limit.  In fact,
it is straightforward to check on how the above solutions reduce to the
conventional Landau wave function.  In this limit of $R\to \infty$, as
we see later, $\llk$ takes a non-negative integer $n$.  As a result,
the confluent hypergeometric function in Eq.~\eqref{eq:Phi1} is
replaced by the Laguerre polynomials through the following
relation~\cite{buchholz2013confluent}:
\begin{equation}
  L_n^l(x) =
  \frac{\Gamma(n+l+1)}{\Gamma(l+1)\,\Gamma(n+1)}\, \hg(-n,l+1,x)
\end{equation}
for any integer $l$, which is simply a definition of the
generalized Laguerre function.

For $l<0$, we cannot use Eq.~\eqref{eq:Phi1} because $\hg(a,b,x)$ is
ill defined for the integer $b\le 0$.  For $l<0$, thus, the above
expression is replaced by
\begin{equation}
  \begin{split}
  \Phi_{l<0} (\lambda,x) &= \frac{(-1)^{-l+1}}{\Gamma(-l+1)} \sqrt{
    \frac{\Gamma(\lambda+1)}{\Gamma(\lambda+l+1)} }\\
    &\quad \times x^{-\frac{l}{2}} e^{-x/2}
    \hg\bigl(-\lambda-l,-l+1,x\bigr)\,.
  \end{split}
  \label{eq:Phi2}
\end{equation}
It should be mentioned that the functions~\eqref{eq:Phi1}
and \eqref{eq:Phi2} reduce to familiar Bessel functions at zero
magnetic field, $B\to 0$, as~\cite{Ebihara:2016fwa}
\begin{equation}
  \begin{split}
    \Phi_l(\llk,\tfrac{1}{2}eB r^2) &\;\;\longrightarrow\;\;
    J_l(\sqrt{2eB \llk}\; r)\,,\\
    \Phi_{l+1}(\llk-1,\tfrac{1}{2}eB r^2) &\;\;\longrightarrow\;\;
    J_{l+1}(\sqrt{2eB \llk}\; r)\,.
  \end{split}
  \label{eq:hg-bessel}
\end{equation}
Also, the negative-energy solutions with the total angular momentum
$j=l+1/2$ are written as
\begin{align}
\label{eq:v+}
  \psi = v_+ &= \frac{e^{i\varepsilon t-ip_zz}}{\sqrt{\varepsilon+m}}
    \begin{pmatrix}
      -i\sqrt{2eB\llk} \philk \\[3pt]
      -p_z \varphilk \\[1pt]
      0 \\
      (\varepsilon+m) \varphilk
    \end{pmatrix} \,,\\
\label{eq:v-}
  \psi = v_- &= \frac{e^{i\varepsilon t-ip_zz}}{\sqrt{\varepsilon+m}}
    \begin{pmatrix}
      -p_z \philk \\[3pt]
      -i\sqrt{2eB\llk} \varphilk \\[3pt]
      -(\varepsilon+m) \philk \\
      0
    \end{pmatrix} \,.
\end{align}
In the Appendix we give the detailed derivation for these
solutions~\eqref{eq:u+}, \eqref{eq:u-}, \eqref{eq:v+},
and \eqref{eq:v-}.  Here, some explanations are needed for the
consistency with Ref.~\cite{Ebihara:2016fwa}, in which we required
that $v_\pm = i\gamma^2 u_\pm^\ast$.  This relation between $u_\pm$ and
$v_\pm$ makes the physical interpretation of antiparticles clear as
long as charge conjugation symmetry $\calC$ is exact.  However, in the
presence of an external $B$, such a naive construction of $v_\pm$
does not satisfy the Dirac equation; under the replacement of
$l\to -l-1$, we see that
$\philk \to \phi_{-l-1,k}$ which would be equal to $\varphilk^\ast$
if $B=0$.  Then, only in the case of $B=0$ does $v_\pm$ in
Eqs.~\eqref{eq:v+} and~\eqref{eq:v-} coincide exactly with the ones
from $v_\pm = i\gamma^2 u_\pm^\ast$ in Ref.~\cite{Ebihara:2016fwa}.
Later, we will return to this point to discuss how to fix $\llk$.

\section{Nondegenerate Landau levels}
\label{sec:Landau}

In finite-size systems, momenta are generally discretized due to the
boundary effect.  As already mentioned in the previous section, we
specifically consider a cylindrical system with the radius $R$ and
assume translational invariance in the longitudinal direction along
the $z$ axis.  In this setup, while $p_z$ is continuous, the
transverse momenta are discretized as a function of $R$.
For scalar fields, for instance, we can impose the Dirichlet boundary
condition at $r=R$, so that we can readily obtain the discretized
momenta~\cite{Davies:1996ks}.  Such a simple treatment is, however,
not applicable to fermionic fields.  This is because Dirac spinors
involve spin-up and spin-down components for which the zeros of the
wave functions appear differently, as is understood in
Eq.~\eqref{eq:phis}.

A possible boundary condition which we will employ here is the
``zero flux constraint'' at $r=R$.  That is, all of the fermionic fluxes
built with $u_\pm$ and $v_\pm$ should be zero at $r=R$, and we express
this condition explicitly as~\cite{Ebihara:2016fwa}
\begin{equation}
  \label{eq:noflux}
  \int_{-\infty}^\infty dz \int_0^{2\pi} d\theta\, 
  \bar \psi \gamma^r \psi\Bigg |_{r=R} = 0\,,
\end{equation}
where we define
$\gamma^r\equiv \gamma^1 \cos\theta + \gamma^2 \sin\theta$.  We note
that Eq.~\eqref{eq:noflux} is not a unique choice but rather that other boundary
conditions for fermionic fields are also possible.  For example, the
MIT-bag-type condition leads to a different type of momentum
discretization, but finite-size effects on fermionic fields are
qualitatively
unchanged~\cite{Ambrus:2015lfr,Chernodub:2016kxh,*Chernodub:2017ref}.

After performing the integration with respect to $\theta$, we see that
the integrand in Eq.~\eqref{eq:noflux} would vanish if
\begin{equation}
\label{eq:PhiPhi}
 \Phi_l(\llk,\alpha) \Phi_{l+1}(\lambda_{l,k'}-1,\alpha) = 0
\end{equation}
for arbitrary $l$, $k$, and $k'$ values.
Here, $\alpha$ is the dimensionless parameter defined by
\begin{equation}
  \label{eq:alpha}
  \alpha \equiv \frac{1}{2} eBR^2 \,.
\end{equation}
Instead of $eB$ or $R$, in this paper, we will frequently refer to
$\alpha$, which is a dimensionless ratio between the magnetic length
$l_B=1/\sqrt{eB}$ and the system size $\ls = R$.  Moreover, this
quantity $\alpha$ corresponds to the conventional Landau degeneracy
factor, i.e.\ $eB(\pi R^2)/(2\pi)$ without boundary distortion.

Now, unlike Ref.~\cite{Ebihara:2016fwa}, the choice of $\llk$ from
Eq.~\eqref{eq:noflux} is not unique;  this nonuniqueness is related
to $v_\pm$, as we mentioned below Eqs.~\eqref{eq:v+} and \eqref{eq:v-}.
In Ref.~\cite{Ebihara:2016fwa} we required
$v_\pm = i\gamma^2 u_\pm^\ast$ from the beginning so that we could keep
charge conjugation symmetry $\calC$.  This symmetry property
gives \textit{another} constraint of invariance under
$l \leftrightarrow -l-1$.  In the present case with $B\neq 0$, there
is no way to keep such symmetry;  nevertheless, it is convenient to
adopt a sufficient condition for Eq.~\eqref{eq:PhiPhi} to be
connected to the $B=0$ limit smoothly, that is,
\begin{equation}
  \begin{split}
  & \Phi_l(\llk,\alpha) = 0 \ \ \qquad\qquad\; \text{for} \quad l\geq 0\,, \\
  & \Phi_{l+1}(\llk-1,\alpha) = 0 \,\qquad \text{for} \quad l\leq -1\,.
  \end{split}
  \label{eq:Phi-condition}
\end{equation}
From the definition of the scalar function $\Phi_l(\lambda,x)$ given
in Eqs.~\eqref{eq:Phi1} and \eqref{eq:Phi2}, we obtain the transverse
momenta discretized as $\plk = \sqrt{2eB\llk}$ with
\begin{equation}
  \llk =
  \begin{cases}
  \xi_{l,k} \,\qquad\qquad\quad \text{for} \quad l\geq 0\,, \\
  \xi_{-l-1,k}-l \qquad \text{for} \quad l\leq -1\,,
  \end{cases}
  \label{eq:lambda}
\end{equation}
where $\xi_{l,k}$ denotes the $k$th zero of $\hg(-\xi,l+1,\alpha)$ as
a function of $\xi$.  We note that $\llk$ depends on $\alpha$;  in
other words, the discretized momenta are functions of the magnetic
field $B$ as well as $R$.

It would be instructive to think of the momentum discretization in the
$B=0$ limit.  From the asymptotic relations~\eqref{eq:hg-bessel}, we
find that the no-flux condition~\eqref{eq:noflux} with
Eq.~\eqref{eq:Phi-condition} leads to the following
discretization:
\begin{equation}
\plk \;\;\underset{eB\to 0}{\longrightarrow}\;\;
\begin{cases}
\displaystyle \zeta_{l,k}/R  \ \, \quad\quad \text{for} \quad l\geq 0\,, \\
\displaystyle \zeta_{-l-1,k}/R \quad \text{for} \quad  l\leq -1\,,
\end{cases}
\label{eq:pperp}
\end{equation}
where $\zeta_{l,k}$ is the $k$th zero of the Bessel function
$J_l(\zeta)$, which matches the preceding
studies~\cite{Ebihara:2016fwa,Ambrus:2015lfr}.  We point out that this
type of momentum discretization respects $\calC$ and $\CP$; i.e.\
wave functions are invariant under
$j\leftrightarrow-j$ (or $l \leftrightarrow -l-1$).

It would be worthwhile to make one more remark about the boundary
condition.  Another boundary condition different from ours can also
lead to the same as Eq.~\eqref{eq:lambda}; namely, one can think of the
following condition~\cite{yoshimasa}:
\begin{equation}
  \int rdr d\theta dz\, \psi_1^\dag \hat H \psi_2 
  = \int rdr d\theta dz\, (\hat H \psi_1)^\dag \psi_2\,,
\end{equation}
where $\hat H \equiv -i\gamma^0\gamma^i (\partial_i + ieA_i)+m\gamma^0$ and $\psi_{1,2}$
are arbitrary solutions of the Dirac equation~\eqref{eq:Deq}.  That
is, the quantized momenta given in Eq.~\eqref{eq:lambda}
[and Eq.~\eqref{eq:pperp} for the $B=0$ case] can result from the
Hermiticity condition for $\hat H$ including the surface term
associated with the integration by parts.

\begin{figure}
\centering
\includegraphics[width=\columnwidth]{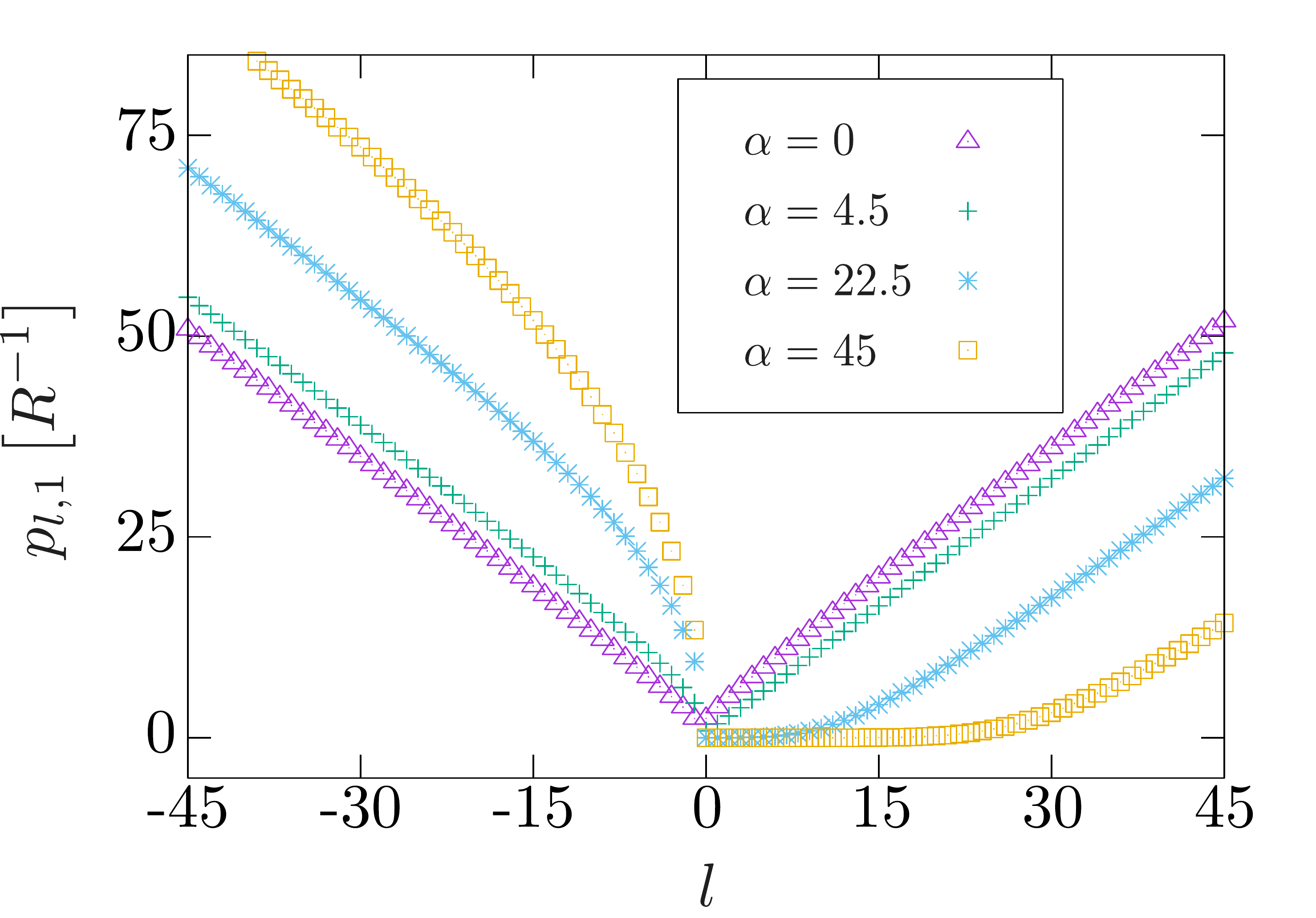}
\caption{%
Lowest transverse momentum $p_{l,1}$ as a function of the angular
momentum $l$ for various $\alpha$'s.}
\label{fig:pl1}
\end{figure}

\begin{figure}
\centering
\includegraphics[width=0.95\columnwidth]{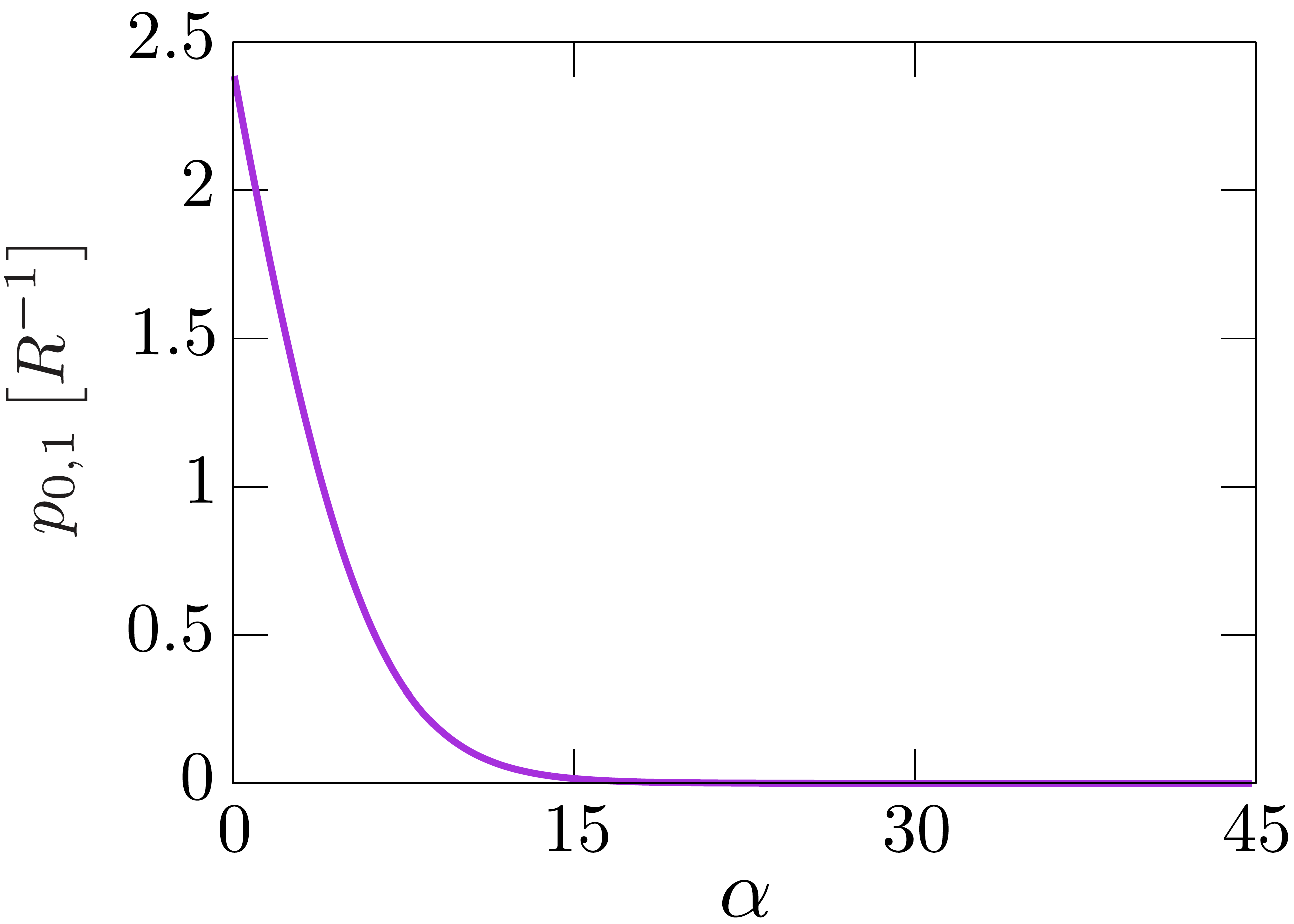}
\caption{%
Lowest momentum $p_{0,1}$ (which gives an energy gap) as a function of
$\alpha=eB R^2/2$.
}
\label{fig:p01}
\end{figure}

In Fig.~\ref{fig:pl1} we plot the lowest transverse momentum $p_{l,1}$
as a function of the angular momentum $l$ for various $\alpha$'s
corresponding to various magnetic fields $B$ or radius $R$ [see
Eq.~\eqref{eq:alpha}].  In the $B=0$ case, as shown by the purple
triangular points in Fig.~\ref{fig:pl1}, positive $l$ modes and
negative $(-l-1)$ modes have a degenerated $p_{l,1}$, which is
immediately understood from the $\CP$ invariance implying
$j \leftrightarrow -j$.  Once a finite magnetic field is turned on,
however, the momenta for the $l>0$ branch are more suppressed than the
$l<0$ branch, as is clear by the green cross, the blue star, and the
magenta square points in Fig.~\ref{fig:pl1}.  Naturally, finite
magnetic fields favor a particular direction of the angular momentum
and break the $\CP$ invariance.  As $\alpha$ increases (i.e., $eB$
or $R$ increases), we see that the lowest momenta become insensitive
to $l$ and the conventional Landau zero modes appear%
~\footnote{More precisely, this is not exactly a gapless mode but rather a ``pseudo'' zero mode, and the exact one is produced in the limit of $\alpha\to\infty$, as seen in Fig.~\ref{fig:p01}.
}.

Figure~\ref{fig:pl1} provides us with more information on the Landau
zero modes peculiar to finite-size systems.  According to the
conventional argument, the Landau degeneracy factor should be given by
$\alpha$, but this is no longer the case for a small $\alpha$;  we
notice in Fig.~\ref{fig:pl1} that $p_{l,1}$ is lifted up from zero at
around $l\simeq 10$ for $\alpha=22.5$.  This means that there are only half of the Landau zero modes, as compared to the conventional
degeneracy factor.  We can intuitively understand this as follows.  The
Landau wave functions with larger $l$'s have a peak position at larger
$r$ due to the centrifugal force, which corresponds to a larger Larmor
radius of the classical cyclotron motion.  The peak width should scale
as $1/\sqrt{eB}$.  For a large enough $\alpha$, the peak is narrow
relative to the system size, and its position hits the boundary at
$r=R$ when $l$ reaches $\simeq \alpha$ (which we have numerically
confirmed for $\alpha=1000$).  For a small $\alpha$, however, the peak
is not well localized, so the Landau zero modes are breached before
$l$ goes up to $\alpha$.  Figure~\ref{fig:pl1} shows a tendency for
the degeneracy of the Landau zero modes to approach $\alpha$ with
increasing $\alpha$;  for $\alpha=45$ the zero modes remain approximately at $l\simeq 30$.

Alternatively, in a slightly different setup of finite-size systems, we can understand the above fact that there emerges less Landau zero modes. 
We suppose that the system is
put not on a cylinder but on a semi-infinite $x-y$ plane with boundary
walls at $x=0$ and $x=\ls$.  If the Landau gauge $A^\mu = (0,-By,0,0)$
is chosen, the peak location of the wave functions is dictated by
$p_x$.  Small momentum modes with $p_x \lesssim \pi/l_B$ or large
momentum modes with $p_x \gtrsim \pi/\ls$ receive strong influences
from the boundary effect.  As a result, for example, a finite-size
graphene ribbon under an external magnetic field has energy dispersion
spectra with large and small $p_x$ modes pushed up, and the Landau
zero modes are seen only for intermediate $p_x$'s~\cite{gusynin2009edge}.

In Fig.~\ref{fig:p01} we show a plot for the lowest momentum
$p_{0,1}$ as a function of $\alpha$.  If there is no boundary,
$p_{0,1}$ must be vanishing.  For a small $\alpha$, however, a finite
gap appears from the boundary effect.  This is obviously so because
$\alpha\to 0$ implies $B\to 0$ and then there is no Landau
quantization.  In this particular limit of $\alpha\to 0$, we find that
$p_{0,1}$ goes to $2.40483/R$, and this value precisely corresponds to
the Bessel zero, $\zeta_{0,1}$, in the discretized
momenta~\eqref{eq:pperp} for $B=0$~\cite{GIORDANO1983221}.  We should emphasize that this
behavior of $p_{0,1}$ is physically quite important.  As argued in
Ref.~\cite{Ebihara:2016fwa}, a rotation alone does not change the
vacuum structure because the induced effective chemical potential
(i.e.\ the rotational energy shift) is
always smaller than the lowest energy gap, $p_{0,1}$.  Once $eB$ becomes bigger than the squared system-size inverse (that is,
$\alpha\gtrsim 10$ from Fig.~\ref{fig:p01}), however, the energy gap
is significantly reduced and the anomalous coupling between the
magnetic field and the rotation is then
manifested~\cite{Chen:2015hfc,Hattori:2016njk}.

\section{Integration Measure and Reweighted wave functions}
\label{sec:measure}

Because the radial momenta are discretized, we replace the transverse
momentum integration with the sums over the quantum numbers, $l$ and $k$,
i.e.
\begin{equation}
  \int \frac{dp_xdp_y}{(2\pi)^2}
  \longrightarrow \frac{1}{\pi R^2} \sum_{l=-\infty}^\infty \sum_{k=1}^\infty 
  \frac{1}{\Nlk^2}\;,
  \label{eq:phase-space}
\end{equation}
where $\Nlk$ represents a weight factor which corresponds to the
integration measure in finite-size systems.  In the $B=0$ case, as
discussed in Ref.~\cite{Ebihara:2016fwa}, the weight factor is deduced
from the Bessel-Fourier expansion, that is, we know that in the limit
of $\alpha\to 0$,
\begin{equation}
 \begin{split}
  \label{eq:Nlkzero}
  \Nlk^2 \;\;\longrightarrow & \;\;
  \frac{2}{R^2}\int_0^R r dr\, \bigl[J_l(p_{l,k}r)\bigr]^2 
 \end{split}
\end{equation}
with the discretized momentum $\plk$ in Eq.~\eqref{eq:pperp}.
From the relation in Eq.~\eqref{eq:hg-bessel}, we extrapolate the
above identification to nonzero $\alpha$ as
\begin{equation}
  \label{eq:Nlk}
  \begin{split}
  \Nlk^2 &= \frac{2}{R^2}\int_0^R r dr\,
  \bigl[\Phi_l(\llk,\tfrac{1}{2}eBr^2)\bigr]^2 \\
  &= \int_0^1 dx\, \bigl[
  \Phi_l(\llk,\alpha x) \bigr]^2\,.
  \end{split}
\end{equation}
We can easily confirm that the $\Nlk$ defined as above satisfies the
asymptotic behavior in Eq.~\eqref{eq:Nlkzero} in the $\alpha\to 0$
limit.  Moreover, we can readily understand that $\Nlk^2$ goes to
$1/\alpha$ in the opposite limit of $\alpha\to \infty$.  Then this
exactly accounts for the appearance of the Landau degeneracy factor,
$\alpha/(\pi R^2)=eB/(2\pi)$, in Eq.~\eqref{eq:phase-space} in the
strong magnetic field limit, which also validates Eq.~\eqref{eq:Nlk}.
Interestingly, as this should be so, we can prove
\begin{equation}
  \int_0^1 dx\,\bigl[\Phi_l(\llk,\alpha x)\bigr]^2
  = \int_0^1 dx\,\bigl[\Phi_{l+1}(\llk-1,\alpha x)\bigr]^2\,.
\end{equation}
This is an important relation;  thanks to this equality, we can
commonly use $\Nlk$ to normalize the four component spinors with both
$\philk$ and $\varphilk$.

As we see in the next section, the propagator involves a spinor matrix
that is a product of two wave functions and, in general, $1/\Nlk^2$
appears together with the propagator.  Thus, the physical meaning of
$\Nlk$ would become more transparent if we define \textit{reweighted}
wave functions by $\Nlk$, i.e.,
\begin{equation}
  \tphilk \equiv \frac{\philk}{\sqrt{\pi R^2} \Nlk}\,,\qquad
  \tvarphilk \equiv \frac{\varphilk}{\sqrt{\pi R^2} \Nlk}
  \label{eq:reweighted}
\end{equation}
for a certain $R$.

Let us explain the interpretation of the reweighted wave functions,
$\tphilk$ and $\tvarphilk$.  We solved the Dirac equation and gave
definitions for $\philk$ and $\varphilk$, but they are not yet
properly normalized, where we simply fixed the overall normalization
to reproduce the conventional expressions in the limit of no boundary
effect.  The important point here is that, for $l>0$, $\varphilk$ may
penetrate outside of $r>R$, while only $\philk$ vanishes at $r=R$;
nevertheless, there is no communication across $r=R$ due to the no-flux
condition.  Therefore, we should normalize the wave functions
within $0\le r\le R$ only.  In other words, we can just presume that
the system is empty for $r>R$;  owing to the no-flux condition, even
in this sharp boundary case, no singularity appears at $r=R$.
To avoid confusion, we must stress that the above description is just
an interpretation, and the denominator in Eq.~\eqref{eq:reweighted}
is, in any case, uniquely fixed in the replacement of the integration with
the discrete sum in Eq.~\eqref{eq:phase-space}.

\begin{figure}
\centering
  \includegraphics[width=0.9\columnwidth]{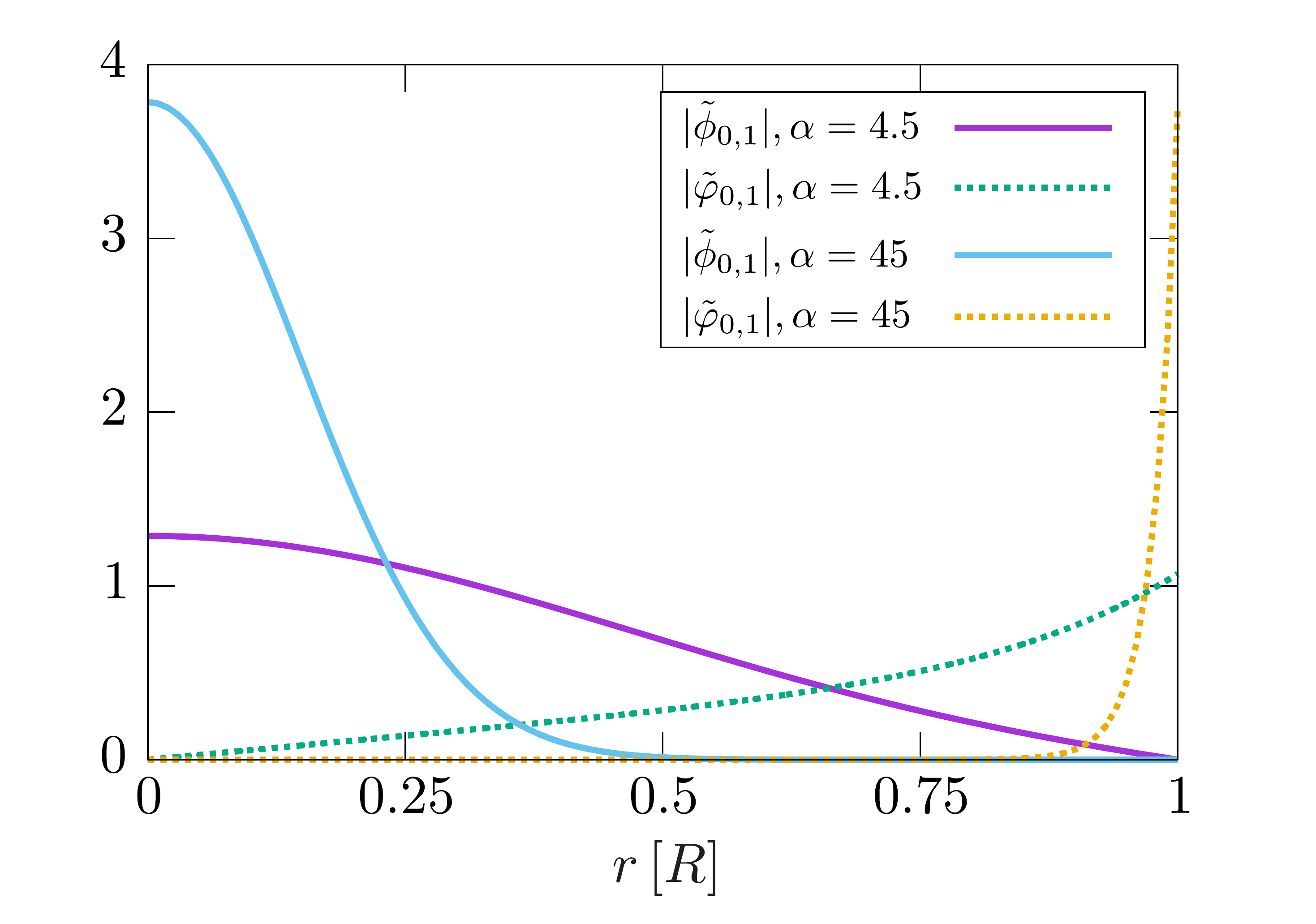}
  \vspace{1em}\\
  \includegraphics[width=0.9\columnwidth]{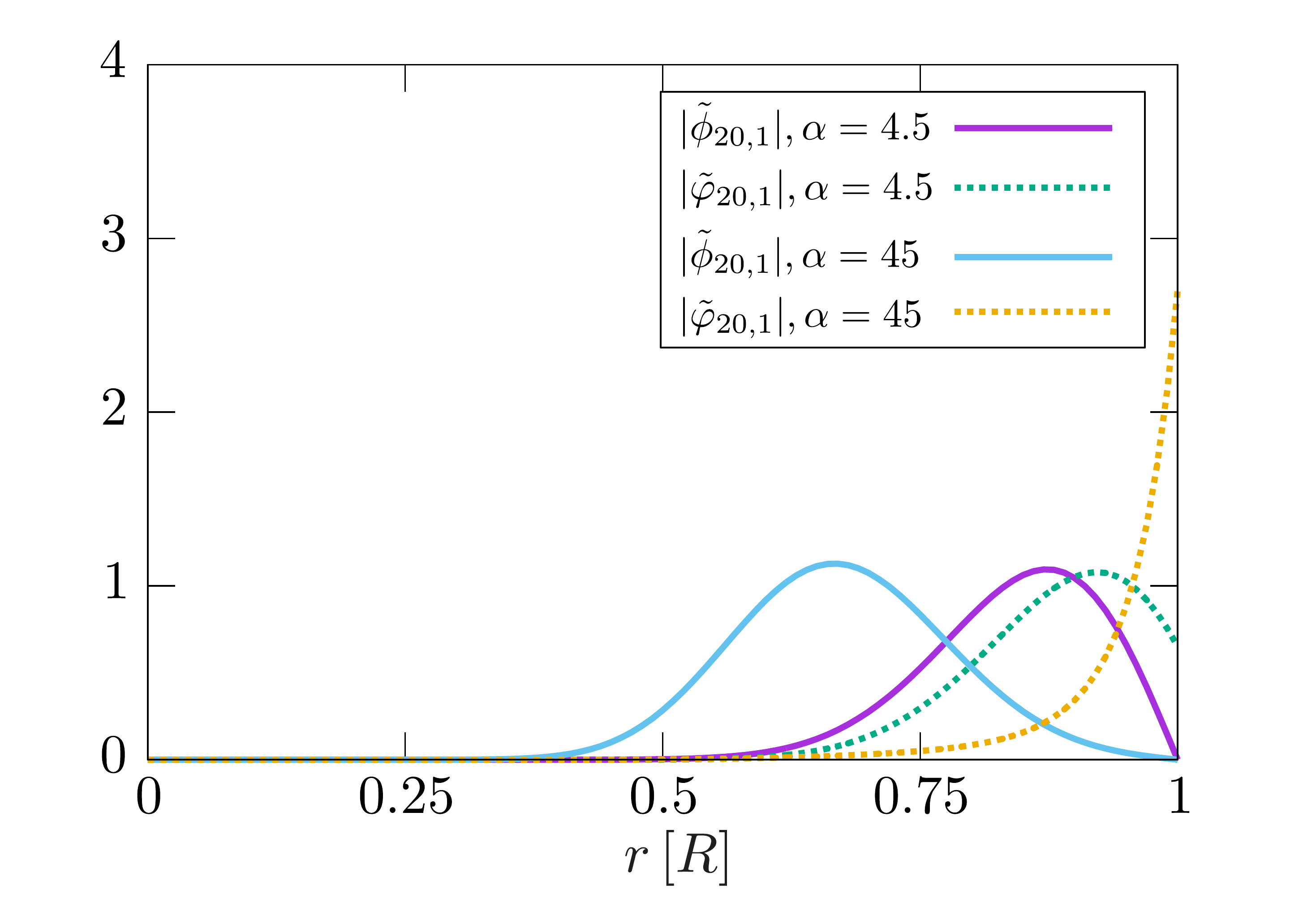}
\caption{%
Radial profiles of $|\widetilde{\phi}_{l,1}|$ (the solid lines) and
$|\widetilde{\varphi}_{l,1}|$ (dashed lines), which are wave functions
normalized by $\sqrt{\pi R^2}N_{l,1}$, where all the quantities are
given in the unit of $R$.  The upper and lower panels show the
wave functions for $l=0$ and $l=20$, respectively.
}
\label{fig:wavefunction}
\end{figure}

From the point of view of a confined picture of wave functions, the
reweighted wave functions, $\tphilk$ and $\tvarphilk$, would make
intuitive sense.  To see the behavior of the reweighted
wave functions, in Fig.~\ref{fig:wavefunction}, we show the radial
dependence of $|\widetilde{\phi}_{l,1}|$ and
$|\widetilde{\varphi}_{l,1}|$ for $l=0$ (upper panel) and $l=20$
(lower panel) (where we chose $k=1$ to see the lowest modes only).
Let us discuss several notable features of the reweighted
wave functions below.

First, we focus on the $l=0$ modes, as depicted in the upper panel of
Fig.~\ref{fig:wavefunction}.  We numerically found
$\xi_{0,1}\simeq 0.041$ for a weak magnetic field ($\alpha=4.5$)
leading to $N_{0,1}^2\simeq 0.19$, and $\xi_{0,1}\simeq 0$ for a
stronger magnetic field ($\alpha=45$) leading to
$N_{0,1}^2\simeq 0.022$.  In fact, if $\xi_{0,1}\simeq 0$, as pointed out
before, $\Nlk^2\simeq 1/\alpha$ is a very good approximation.

Because $l=0$ corresponds to the $S$-wave, $|\phi_{0,1}|$ is centered
around $r=0$ and becomes more localized for larger $\alpha$'s.  As
noticed in Eq.~\eqref{eq:phis}, on the other hand, $|\varphi_{0,1}|$
has a $l=1$ component of the $P$-wave, so the wave function 
peaks near the boundary due to the centrifugal force.  It is an
interesting observation that $|\varphi_{0,1}|$ gets more and more
sharply attached to the boundary with increasing $\alpha$.  In the
infinite size limit $R\to \infty$ (i.e., $\alpha\to \infty$), there is
no contribution at all from $|\varphi_{0,1}|$, which means that
both $u_\pm$ are eigenstates of the spin $s_z=\frac{1}{2}\sigma_z$ with an
eigenvalue $+\frac{1}{2}$, that is, spin-up states.  This observation
is consistent with the fact that the Landau zero modes have only one
spin state.

Next, we consider the $l=20$ modes by looking at the lower panel in
Fig.~\ref{fig:wavefunction}.  The behavior is qualitatively different
from the $l=0$ case.  The most nontrivial point is seen in the
difference between $|\widetilde{\varphi}_{0,1}|$ in the upper panel
and $|\widetilde{\varphi}_{20,1}|$ in the lower panel for
$\alpha=4.5$.  As explained above, the centrifugal force with $l=1$
pushes $|\widetilde{\varphi}_{0,1}|$ toward $r=R$, and one would
expect that such centrifugal effects must be greater for $l=20$.
However, $|\widetilde{\phi}_{20,1}|$ is centered rather away from
$r=R$, which seems to be quite counterintuitive.  We can resolve this
puzzle from an indirect constraint from $|\widetilde{\phi}_{l,1}|$;
for the $l=0$ case, $|\widetilde{\phi}_{0,1}|$ is not modified much by
the boundary because the wave function tail at $r=R$ is negligibly
small from the beginning.  However, for the $l=20$ case,
$|\widetilde{\phi}_{20,1}|$ is significantly distorted and this
boundary effect is strong enough to distort
$|\widetilde{\varphi}_{20,1}|$ as well.

Another interesting observation for the $l=20$ wave functions is that
the spin-up and the spin-down states are not really separable unlike
the $l=0$ case.  We recall that in the upper panel of
Fig.~\ref{fig:wavefunction} the region with $r<R$ is dominated by
$|\widetilde{\phi}_{0,1}|$ only and the wave function inevitably
becomes the spin-up eigenstate.  In the $l=20$ case, however, due to
the centrifugal force, all of the wave functions are shifted in the
vicinity of the boundary, and there, $|\widetilde{\phi}_{20,1}|$ and
$|\widetilde{\varphi}_{20,1}|$ always coexist;  in other words, the
Landau degeneracy is violated for large $l$'s, as we already saw in
Fig.~\ref{fig:p01}.

We emphasize the importance of the wave function behavior around
$r=R$.  In this way the wave functions at larger $l$'s are accumulated
near $r=R$ and the low-energy dynamics closer to the boundary is more
prominently affected by the magnetic background.  For instance, as we
will confirm in the next section, the dynamical mass enhancement by
the magnetic field is further strengthened near the boundary.  As a
side remark, we note that the mode accumulation around the boundary
has no contradiction with the Pauli exclusion principle because all
accumulated modes are labeled by different quantum numbers.

\section{Boundary enhancement of the magnetic catalysis}

We will proceed to some concrete calculations to demonstrate the
interplay between the magnetic and the surface effects.  We will
estimate the dynamical mass in the local density approximation using
an NJL model.  The qualitative features are robust in the sense that it does not depend on a model choice, however, as is clear from the
physical discussions in the previous section.

The most fundamental ingredient for concrete calculations is the
propagator, $S$, which can be constructed from the
solutions~\eqref{eq:u+}, \eqref{eq:u-}, \eqref{eq:v+},
and~\eqref{eq:v-}.  Then, in terms of the Dirac indices, $S$ is a
$4\times 4$ matrix whose form is given by
\begin{equation}
  \begin{split}
    & S^{\alpha\beta}(x,x') 
    = \ i \int \frac{dp^0dp_z}{(2\pi)^2}
      \sum_{l=-\infty}^\infty\sum_{k=1}^\infty
      \frac{1}{\pi R^2 \Nlk^2}\\
     &\qquad \times
      \frac{e^{-i p^0 (t-t') +ip_z (z-z')}}
      {(p^0)^2-\varepsilon^2+i\epsilon}
    \mathcal{S}_{l,k}^{\,\alpha\beta}(p;r,\theta,r',\theta')\,,
  \end{split}
  \label{eq:propagator}
\end{equation}
where the spinor matrix $\mathcal{S}_{l,k}^{\,\alpha\beta}$ in the
Dirac representation reads
\begin{equation}
  \mathcal{S}_{l,k}(p;r,\theta,r',\theta') =
    \begin{pmatrix}
      \mathcal{M}_{l,k}^{(+)} & \mathcal{N}_{l,k}^{(+)} \\
      \mathcal{N}_{l,k}^{(-)} & \mathcal{M}_{l,k}^{(-)}
  \end{pmatrix}
\end{equation}
with
\begin{equation}
\begin{split}
  \mathcal{M}_{l,k}^{(\pm)} &\equiv 
    \begin{pmatrix}
      (\pm p_0 + m) \phi_{l,k} \phi'_{l,k} & 0 \\
      0  & (\pm p_0 + m)  \varphi_{l,k} \varphi'_{l,k}
    \end{pmatrix}\,, \\
  \mathcal{N}_{l,k}^{(\pm)} &\equiv 
    \begin{pmatrix}
      -p_z \phi_{l,k} \phi'_{l,k}
      & \pm i\sqrt{2eB\llk}\phi_{l,k} \varphi'_{l,k} \\
      \mp i\sqrt{2eB\llk} \varphi_{l,k} \phi'_{l,k}
      & p_z \varphi_{l,k} \varphi'_{l,k}
    \end{pmatrix}\,.
\end{split}
\end{equation}
In the above equations, we use a short notation for the wave functions;
$\philk=\philk(r,\theta)$,
$\phi'_{l,k}=\philk^*(r',\theta')$,
$\varphilk=\varphilk(r,\theta)$, and
$\varphi'_{l,k}=\varphilk^*(r',\theta')$.  We note that
$\pi R^2 \Nlk^2$ in Eq.~\eqref{eq:propagator} may have been absorbed
into redefinition of $\philk\to\tphilk$ and $\varphilk\to\tvarphilk$.

To study the boundary effect for the dynamical mass generation
associated with the spontaneous breaking of chiral symmetry, we
analyze the NJL model whose Lagrangian density is
\begin{equation}
  \Lag
  = \bar{\psi} \, i\gamma^\mu
    (\partial_\mu+ieA_\mu) \psi
    + \frac{G}{2}\bigl[ (\bar{\psi}\psi)^2
    + (\bar{\psi}i\gamma_5\psi)^2 \bigr]\;.
  \label{eq:NJL}
\end{equation}
In the mean-field approximation (which is justified when there are
infinitely many fermion species), the gap equation or the condition to
minimize the thermodynamic potential is written as
\begin{equation}
  m = G\, \tr[S(x,x)]\,.
\label{eq:gapeq0}
\end{equation}
Since translation invariance is lost along the radial direction, the
dynamical mass has the $r$ dependence, and thus we should regard
Eq.~\eqref{eq:gapeq0} as a functional equation to determine a function
$m(r)$.  It is, however, numerically demanding to solve this
functional equation self-consistently.  Besides, our present purpose
is not to quantify the effects but to demonstrate robust features of
the surface effects.  Thus, we reasonably simplify the problem by
employing the local density approximation under an assumption of
$|\partial_r m(r)| \ll m(r)^2$~\cite{Jiang:2016wvv}.  Then, we can
approximately treat the energy dispersion relation as simple as
$\varepsilon(r) = \sqrt{2eB\llk+ p_z^2 + m(r)^2}$.
Utilizing Eq.~\eqref{eq:propagator} and inserting a ultraviolet
regulator, we write the explicit form of the gap equation as
\begin{equation}
  \begin{split}
    \frac{m(r)}{G} 
    & = m(r) \int_{-\infty}^\infty \frac{dp_z}{2\pi}
      \sum_{l=-\infty}^\infty\sum_{k=1}^\infty
      \frac{f(p\,;\Lambda,\delta\Lambda)}{\pi R^2 \Nlk^2}\\
      & \times 
      \frac{\bigl[\Phi_l(\llk,\tfrac{1}{2}eBr^2)\bigr]^2 
      + \bigl[\Phi_{l+1}(\llk-1,\tfrac{1}{2}eBr^2)\bigr]^2}  
        {\varepsilon(r)} 
       \,.
  \end{split}
  \label{eq:gapeq}
\end{equation}
Here, we note that the choice of the ultraviolet regulator is a part
of the model definition, and in our numerical calculations presented
below, we adopt a smooth 3-momentum cutoff function as
follows~\cite{Chen:2015hfc}:
\begin{equation}
  f(p\,;\Lambda,\delta\Lambda) = \frac{\sinh(\Lambda/\delta\Lambda)}
    {\cosh(p/\delta\Lambda)
     + \cosh(\Lambda/\delta\Lambda)}
\label{eq:cutoff}
\end{equation}
with $p=\sqrt{2eB\llk + p_z^2}$.  To discuss the magnetic catalysis,
the proper-time regularization~\cite{Schwinger:1951nm} and the
Pauli-Villars regularization would be a common choice in NJL model
studies (see e.g., see Ref.~\cite{Klimenko:1991he,Gusynin:1994re}).  It
is, however, known that a naive momentum cutoff with a step function
could also give a qualitatively correct result, as long as the smearing
parameter $\delta\Lambda$ is not too small~\cite{Gorbar:2011ya}.
Therefore, the above simple $f(p;\Lambda,\delta\Lambda)$  value should
suffice for our present purpose of qualitative analysis.   For the
numerical calculation we chose the model parameters as
\begin{equation}
  G = 24\,\Lambda^{-2}\,,\qquad \delta\Lambda = 0.05\,\Lambda\,.
\end{equation}
Here, we can trivially scale out $\Lambda$ by measuring all of the
quantities in units of $\Lambda$.  
Also we fix the system size to be
\begin{equation}
  R = 30\,\Lambda^{-1}\,.
\end{equation}
This value itself is not relevant for our discussions.  For
$\Lambda\simeq 1\,\text{GeV}$ (that is, a QCD scale), the above choice
of the system size $R=30\,\Lambda^{-1}$ corresponds to the typical
radial scale of heavy ions, $R \simeq 6\,\text{fm}$.
In units of $\Lambda$, in this model with $B=0$ and $R\to\infty$, the critical coupling is
\begin{equation}
  G_c = 19.58\,\Lambda^{-2}\,,
\end{equation}
which is smaller than the present $G$.

\begin{figure}
  \centering
  \includegraphics[width=0.9\columnwidth]{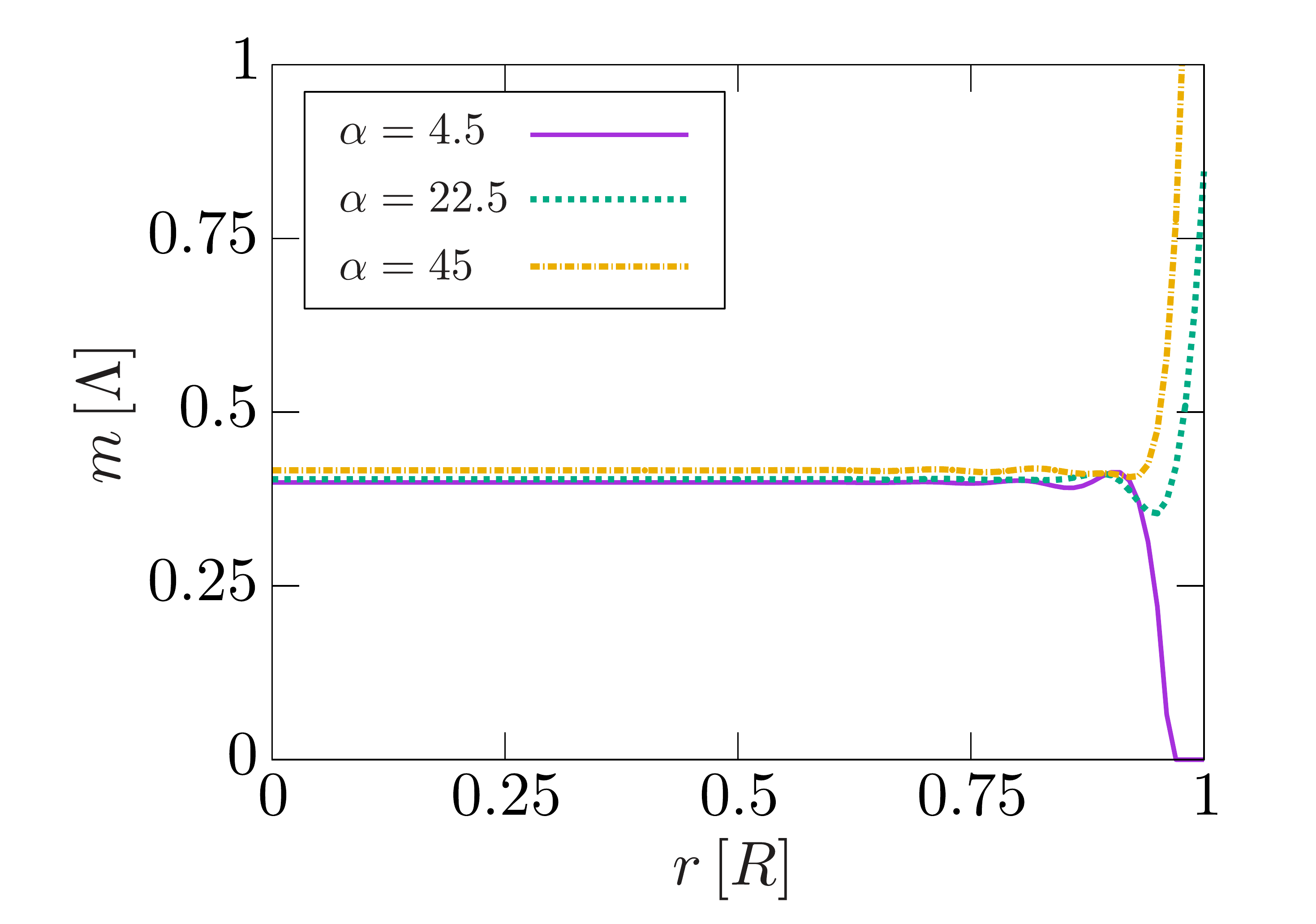}
\caption{%
Dynamical mass as a function of the radial coordinate $r$ for the
choice of $R = 30 \Lambda^{-1}$.  Near the boundary the dynamical mass
rapidly increases due to the accumulation of the boundary modes.
}
\label{fig:mass}
\end{figure}

In Fig.~\ref{fig:mass} we show the dynamical mass $m(r)$ solved from
the gap equation in the local density approximation.  We see from
Fig.~\ref{fig:mass} that the magnetic field effect is minor for
$\alpha=4.5$.  The $r$ dependence of the dynamical mass is flat up to
$r\simeq 0.7R$, and $m(r)$ becomes oscillatory for
$r\gtrsim 0.7R$.  Such oscillation results from the boundary effect,
and its exact form depends on the regularization
$f(p;\Lambda,\delta\Lambda)$ as well as the system size.  Actually,
for a larger $R$, the discretized momentum spacing is smaller (which is
$\propto 1/R$), and thus the oscillating period should be smaller
accordingly.  For the even larger $r\simeq R$, the dynamical
mass eventually vanishes.  This oscillating and vanishing behavior of $m(r)$ is
quite similar to what is observed in the $B=0$ case (see Fig.~1 in
Ref.~\cite{Ebihara:2016fwa}).  We also comment on the validity
of the local density approximation.  The required condition,
$|\partial_r m(r)|\ll m(r)^2$ is satisfied for almost all $r$'s except
the region very close to $R$.

In contrast to $\alpha=4.5$, the dynamical mass behavior for stronger
magnetic fields ($\alpha = 22.5$ and $45$ in Fig.~\ref{fig:mass}) is
qualitatively different.  As long as $r$ is away from the boundary, a
flat plateau continues, until oscillations appear around $r\simeq 0.7R$.
Then, $m(r)$ does not vanish but is pushed up as $r$ approaches $R$.
This abnormally enhanced magnetic catalysis (called the
surface magnetic catalysis in this work) is a consequence
of the interplay between the magnetic field and the boundary effect.

We shall explain how to understand the surface magnetic catalysis in
terms of the wave functions.  We have already seen that the spin-down
mixture by $\tvarphilk$ piles up near $r\simeq R$ for a large $l$, as
shown in the lower panel of Fig.~\ref{fig:wavefunction}.  If there
were no boundary, the peak position of the wave functions with large
$l$ would be at a far distance.  However, in the presence of the boundary
at $r=R$, these modes, which would have no contribution without
a boundary, come to make a finite contribution near $r\simeq R$.  Then, the
gap equation~\eqref{eq:gapeq} receives a contribution of spin-down
boundary modes with various $l$'s.  We could say, in other words, that
the surface magnetic catalysis is induced by the combination of the
incomplete spin alignment of the Landau levels seen in
Sec.~\ref{sec:Landau} and the reweighting factor from the integration
measure argued for in Sec.~\ref{sec:measure}.

\section{Conclusion}
\label{sec:conclusion}

In this paper, imposing a proper boundary condition in terms of the
fermionic flux (the same conclusion can be drawn from the
Hermiticity of the Hamiltonian), we analyzed the finite-size effect on
fermionic matter coupled with an external magnetic field.  We
obtained incomplete or nondegenerate Landau levels; that is, for
states with larger angular momenta relative to the system size, the
Landau quantized spectra are not degenerate.  Also, we noticed that the
spin-up and spin-down structures of the wave functions are
significantly changed by the finite-size effect.  In the thermodynamic
limit of infinite volume, only the spin-up modes (if the magnetic field
is positive along the quantum axis of the angular momentum) occupy the
Landau zero modes, and the spin-down modes become irrelevant because
the spin-down modes are tightly localized in the vicinity of the
boundary at an infinitely great distance.  In finite-size systems, however,
the magnetic field partially overcomes this spin separation and forms
the gapless Landau zero modes for both spin-up and spin-down states.
This pairwise structure of spin-up modes in the bulk and spin-down
modes at the surface is quite remarkable for magnetic phenomena related to
chirality imbalance.  For instance, in finite-size systems, even
though an anomalous fermionic current density is nonzero in bulk, the
whole current would vanish together with the surface
contribution~\cite{kamata}.

In this paper, we found a novel aspect of the magnetic
catalysis peculiar to finite-size systems;  the catalyzing effect on
the dynamical mass is more intense in the vicinity of the boundary,
which is called the surface magnetic catalysis in this work.
Because the surface magnetic catalysis shows a sharp enhancement of
the dynamical mass at the surface, strictly speaking, we must say that the
local density approximation, in which spatial derivatives of the
dynamical mass are neglected, adopted in the present work might be not
reliable enough.  We must stress, however, that the origin of such a
strong enhancement can be explained by the accumulation of many
spin-down zero modes near the boundary, which does not rely on
any model or approximation.  Therefore, even including the higher
order derivative terms of the dynamical mass, nothing qualitative
should be changed.  Regardless of the model or the approximation,
similar spatial profile of the dynamical mass or the condensate must
be reproduced.  Furthermore, we note that the geometrical shape of the
boundary is not relevant to the accumulation of the low-energy modes.
Hence, lattice numerical simulations could test the surface magnetic 
catalysis in a realistic finite-size system (for instance, graphene~\cite{DeTar:2016vhr,*DeTar:2016dmj}) if not the periodic boundary 
condition but an appropriate no-flux boundary condition is formulated 
in terms of the link variables.

The findings in this paper have various applications.  For
Dirac and Weyl semimetals we expect fruitful phase structures from the
proper treatment of the finite-size effect.  In
Ref.~\cite{Gorbar:2013qsa}, the authors argue that an external
magnetic field leads to the dynamical transformation from a Dirac
semimetal (a state without the chiral shift~\cite{Gorbar:2009bm}) to a
Weyl semimetal (a state with the chiral shift).  This is the case for
large systems.  According to our result, the magnetic property of
the boundary should differ from that of the bulk, and thus it would
be intriguing to revisit the possibility of the dynamical
transformation including the surface effect.

Another interesting extension is the coupling to rotation.
For instance, the anomalous coupling with the magnetic field and the
rotation~\cite{Ebihara:2016fwa,Hattori:2016njk} should lead to a
fascinating effect on the energy-momentum tensor of a quark-gluon
plasma~\cite{Hernandez:2017mch,mameda}.  Besides, the
interplay between the magnetic field and the rotation should influence
the dynamical symmetry breaking and the equation of state.  As
discussed in Ref.~\cite{Chen:2015hfc}, in rotating matter, the
magnetic catalysis and the inverse magnetic catalysis are driven,
respectively, by small and large rotational effects.  At a short distance $r$ from the rotational center, the magnetic
catalysis is realized because the centrifugal force,
which is proportional to $r$, is still small.  Since the edge region near the boundary
is heavily affected by the magnetic field, on the other hand, it is
nontrivial whether the inverse magnetic catalysis really takes place
around the boundary once the results in Ref.~\cite{Chen:2015hfc} are
augmented by the finite-size effects.  The quantitative details of
the chiral structure in magnetized rotating systems deserve
further investigations, and we will report our progress in
forthcoming papers.

\acknowledgments
The authors thank Yoshimasa~Hidaka for the useful discussions.
K.~M. thanks Toru~Kojo and Igor~Shovkovy for the valuable comments.
H.-L.~C. and X.-G.~H. are supported by the Young 1000 Talents Program of China, 
NSFC through Grants No.~11535012 and No.~11675041, and the Scientific Research Foundation 
of State Education Ministry for Returned Scholars.
K.~F. was supported by Japan Society
for the Promotion of Science (JSPS) KAKENHI Grants No.~15H03652 and~15K13479.
K.~M. was partially supported by Grant-in-Aid for JSPS Fellows, Grant No.~15J05165.

\begin{appendix}

\section{Solving the Dirac equation}
\label{app:Deq}

We derive the solutions~\eqref{eq:u+}, \eqref{eq:u-}, \eqref{eq:v+},
and~\eqref{eq:v-}.  The Dirac equation for fermions confined in a
finite-size system under an external magnetic field is given by
\begin{equation}
  (i\gamma^\mu D_\mu -m)\psi = 0\,,
\end{equation}
where $D_\mu = \partial_\mu + ieA_\mu$ is the covariant derivative
with the symmetric gauge $A^\mu = (0,-By/2,Bx/2,0)$.  Multiplying
$(i\gamma^\nu D_\nu +m)$ by the above Dirac equation and changing to
the cylindrical coordinates, we can rewrite Eq.~\eqref{eq:Deq} as
follows:
\begin{equation}
  \begin{split}
  & \Biggl[-\partial_t^2 + \partial_z^2 -m^2 
  + \partial_r^2 + \frac{1}{r}\partial_r +\frac{1}{r^2}\partial_\theta^2\\
  &\qquad + eB(-i\partial_\theta+\sigma^{12}) - \biggl(\frac{eBr}{2}\biggr)^2\
  \Biggr]\psi = 0\,.
  \label{eq:Deq-cylinder}
  \end{split}
\end{equation}
with $\sigma^{12} = \tfrac{i}{2}[\gamma^1,\gamma^2] = \text{diag}(\sigma_z,\sigma_z)$.
Since the $t$- and $z$-dependent terms are separately solved in the form
of plane waves, we can parametrize two linear independent solutions
with positive energy as
\begin{equation}
  \psi = u_{\pm} =  e^{-i\varepsilon t+ip_z z}
  \begin{pmatrix}
  f_{1\pm} (r,\theta)\\
  f_{2\pm} (r,\theta)
  \end{pmatrix} \,,
\end{equation}
where $\pm$ refers to different polarizations.

Let us first focus on $f_{1\pm}$.  While the total angular momentum,
$\hat J_z = \hat L_z + \hat S_z = -i\partial_\theta + \tfrac{1}{2}\sigma^{12}$,
is a good quantum number in the present system, neither $\hat L_z$ nor
$\hat S_z$ is.  For this reason it is convenient to choose $u_\pm$ as
an eigenstate of $\hat J_z$ with its common eigenvalue denoted by $j$.
We here employ the Dirac representation for $\gamma^\mu$'s, i.e.,
\begin{equation}
  \gamma^0 =
  \begin{pmatrix}
  1 & 0 \\
  0 & -1
  \end{pmatrix} \,, \qquad
  \gamma^i = 
  \begin{pmatrix}
  0 & \sigma^i \\
  -\sigma^i & 0
  \end{pmatrix} \,.
\end{equation}
Then, we fix the angular part of the two component function $f_{1\pm}$
as
\begin{equation}
  f_{1\pm} (r,\theta) = e^{il_\pm \theta} \tilde f_{1\pm}(r) \chi_\pm
\end{equation}
with $\sigma_z \chi_\pm = \pm \chi_\pm$ and $l_+ +1/2 = l_- -1/2 = j$.
From Eq.~\eqref{eq:Deq-cylinder} we find the equation of motion for
the radial part, $\tilde f_{1\pm}$, which reads
\begin{equation}
\begin{split}
 &\Biggl[ \partial_r^2 + \frac{1}{r}\partial_r - \frac{l_\pm^2}{r^2} + 2eB\llk \\
 & \qquad\qquad + eB(l_\pm \pm 1)  - \biggl(\frac{eBr}{2}\biggr)^2\ \Biggr] \tilde f_{1\pm} = 0
\end{split}
\label{eq:eos_tilf}
\end{equation}
with the dispersion relation,
\begin{equation}
  2eB\llk = \varepsilon^2-p_z^2-m^2\,.
\end{equation}
Using the scalar function $\Phi_l(\lambda,\tfrac{1}{2}eBr^2)$ defined in
Eqs.~\eqref{eq:Phi1} and \eqref{eq:Phi2}, we identify the solutions
for this equation as $\tilde f_{1+}=\Phi_l(\llk,\tfrac{1}{2}eBr^2)$ and
$\tilde f_{1-}=\Phi_{l+1}(\llk-1,\tfrac{1}{2}eBr^2)$,
where we introduce the quantum number for $\hat L_z$,
\begin{equation}
 l\equiv j-1/2\,, \quad \text{i.e.\ } \quad l=l_+ = l_--1\,.
\end{equation}
Thus, we find that $f_{1\pm}$ is represented as follows:
\begin{equation}
  f_{1+} (r,\theta) = \philk\chi_+\,,\qquad
  f_{1-} (r,\theta) = \varphilk \chi_- \,,
\end{equation}
with $\philk$ and $\varphilk$ in Eq.~\eqref{eq:phis}.

Also, we can solve the lower components $f_{2\pm}$ from
\begin{equation}
  (\varepsilon+m) f_{2\pm} = (-i\bsigmap \cdot \bDp + \sigma_z p_z) f_{1\pm}\,,
\label{eq:f12}
\end{equation}
which follows from the Dirac equation in the Dirac representation.  In
the cylindrical coordinates, the covariant derivative term,
$-i\bsigmap\cdot\bDp$, is represented as
\begin{equation}
-i\bsigmap\cdot\bDp
 =
\begin{pmatrix}
 0 & a^\dag \\
 a & 0
\end{pmatrix}\,,
\end{equation}
where we introduce the ladder operators defined by
\begin{equation}
\begin{split}
a &\equiv 
-i e^{i\theta} (\partial_r + i r^{-1}\partial_\theta + eBr/2)\,,\\
a^\dag &\equiv 
-i e^{-i\theta} (\partial_r - i r^{-1}\partial_\theta - eBr/2)\,.\\
\end{split}
\end{equation}
In fact, we can explicitly check to see that $a$ and $a^\dag$ act as the
ladder operator on $\philk$ and $\varphilk$:
\begin{equation}
\begin{split}
& a\philk = i\sqrt{2eB\llk} \varphilk\,, \\
& a^\dag \varphilk = -i\sqrt{2eB\llk} \philk\,.
\end{split}
\end{equation}
From these relations and the explicit form of $\tilde f_{1\pm}$, we can
solve Eq.~\eqref{eq:f12} for $f_{2\pm}$ as
\begin{equation}
\begin{split}
 f_{2+} & 
= \frac{p_z}{\varepsilon+m} \philk\chi_+ 
+ \frac{i\sqrt{2eB\llk}}{\varepsilon+m} \varphilk\chi_- \,,\\
 f_{2-}
& = \frac{-i\sqrt{2eB\llk}}{\varepsilon+m} \philk\chi_+ 
+ \frac{-p_z}{\varepsilon+m} \varphilk\chi_-\,,
\end{split}
\end{equation}
which finally amounts to Eqs.~\eqref{eq:u+} and \eqref{eq:u-} for the
positive-energy solution.

In the same way, we find the negative-energy solution, $v_{\pm}$.  We
suppose that the solution takes the following form:
\begin{equation}
\psi = v_{\pm} =  e^{i\varepsilon t-ip_z z}
\begin{pmatrix}
g_{1\mp} (r,\theta)\\
g_{2\mp} (r,\theta)
\end{pmatrix}
\end{equation}
with
\begin{equation}
 g_{2+} (r,\theta)
  = \varphilk \chi_+ \,,\qquad
 g_{2-} (r,\theta) 
  = -\philk \chi_- \,.
\end{equation}
Then, the Dirac equation fixes the form of the upper component
$g_{1\pm}$ through
\begin{equation}
(\varepsilon+m) g_{1\pm} = (i\bsigmap \cdot \bDp - \sigma_z p_z) g_{2\pm}\,.
\label{eq:g12}
\end{equation}
We can explicitly solve this equation, leading to
\begin{equation}
\begin{split}
 g_{1+} & 
= \frac{-p_z}{\varepsilon+m} \philk\chi_+ 
 +\frac{-i\sqrt{2eB\llk}}{\varepsilon+m} \varphilk\chi_- \,,\\
 g_{1-}
& = \frac{-i\sqrt{2eB\llk}}{\varepsilon+m} \philk\chi_+ 
+ \frac{-p_z}{\varepsilon+m} \varphilk\chi_-\,.
\end{split}
\end{equation}
Hence, we obtain Eqs.~\eqref{eq:v+} and \eqref{eq:v-}.
\end{appendix}

\bibliography{rotation}
\end{document}